\documentclass[sigconf,nonacm]{acmart}

\setcopyright{none}
\usepackage{graphicx}
\usepackage{mathtools}
\usepackage{bm}
\usepackage{algorithm}
\usepackage{algorithmic}
\usepackage{booktabs}
\usepackage{multirow}
\usepackage{makecell}
\usepackage{array}
\usepackage{tabularx}

\title{\textsc{Forge}: An Agentic System for ERGM Specification and Interpretation in Social Networks}
\author{Yidan Sun}
\email{yidans@usc.edu}
\affiliation{%
  \institution{University of Southern California}
  \city{Los Angeles}
  \state{California}
  \country{USA}
}
\author{Mayank Kejriwal}
\email{kejriwal@isi.edu}
\affiliation{%
  \institution{Information Sciences Institute, University of Southern California}
  \city{Marina del Rey}
  \state{California}
  \country{USA}
}

\hypersetup{pdfauthor={Yidan Sun and Mayank Kejriwal}}

\begin{document}

\begin{abstract}
Exponential Random Graph Models (ERGMs) provide a statistical framework for testing hypotheses about how ties form using network structure and node attributes. Their usefulness depends on model specification, which often requires researchers to repeatedly propose models, fit them, check how well they reproduce the observed network, and revise them when they do not. We present \textsc{Forge} (Formation-Oriented Reasoning with Guarded ERGMs), an agentic system that performs this process using a language model and external statistical tools. Given an observed network and a domain description, \textsc{Forge} proposes hypotheses about tie formation, translates them into candidate ERGM specifications, evaluates them statistically, refines the best candidate using goodness-of-fit feedback, and returns a human-readable interpretation of the final model. On 16 benchmark networks, \textsc{Forge} improves on the best baseline on 13 networks and matches it on the other three, with a median 24.8\% improvement in goodness of fit. On 60 controlled networks with known generating specifications, \textsc{Forge} achieves exact specification recovery in 78.3\% of cases and attains 90.8\% term-level F1. These results show that \textsc{Forge} can reduce the manual trial-and-error involved in ERGM specification while making the process more reproducible and keeping model selection grounded in statistical evidence.
\end{abstract}



\maketitle
\pagestyle{plain}

\begin{figure*}[t]
\centering
\includegraphics[width=0.95\linewidth]{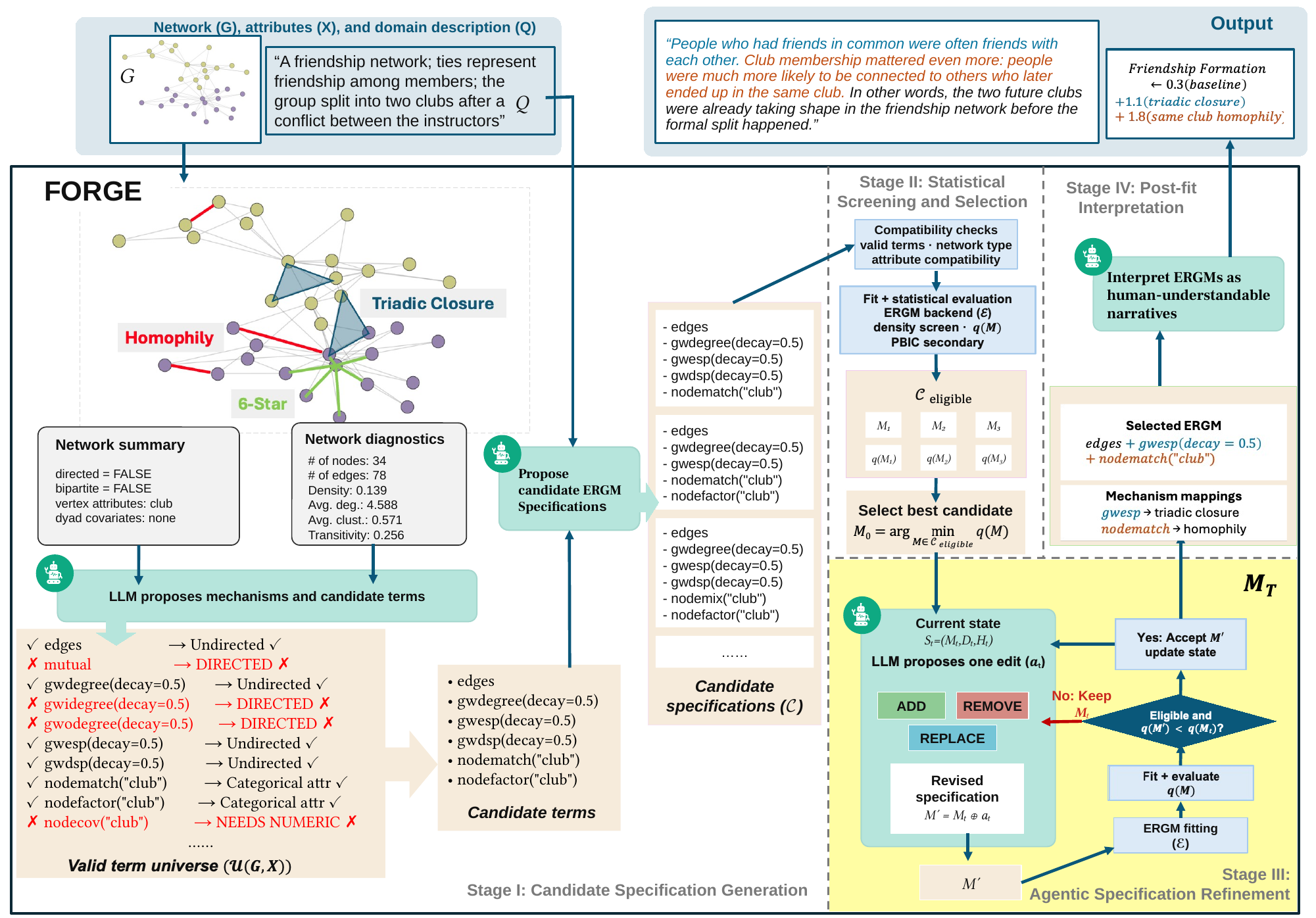}
\caption{
The \textsc{Forge} workflow:
Stage~I generates candidate specifications from network information;
Stage~II applies statistical screening and selects the candidate that passes the statistical checks and has the lowest goodness-of-fit (GOF) discrepancy $q(M)$;
Stage~III proposes one local edit per round and keeps it only when the revised specification passes the same checks and lowers $q(M)$;
Stage~IV interprets the fixed final ERGM in terms of formation mechanisms.
}
\Description{Four-stage Forge workflow: mechanism-guided candidate generation; compatibility screening, ERGM estimation, post-fit eligibility determination, and model selection; one-edit-at-a-time diagnostic refinement; and post-fit mechanism interpretation.}
\label{fig:forge-framework}
\end{figure*}

\section{Introduction}

Web and social systems form networks through replies, mentions, reposts, communication, and collaboration \cite{sousa2010replies,kwak2010twitter,borgatti2009network,newman2003structure}. Understanding why these ties form is a central problem in network science. Ties may reflect reciprocity, triadic closure, shared attributes, or homophily \cite{snijders2010introduction,lusher2013exponential}. Yet these processes are not directly observed, and similar network structures can arise for different reasons. High clustering, for example, may reflect triadic closure, shared interests, or participation in the same activity. The challenge is therefore not simply to describe the network, but to determine which explanations of tie formation are supported by the observed data.

Several statistical frameworks are used to study network formation, including stochastic block models, latent-space models, stochastic actor-oriented models, and Exponential Random Graph Models (ERGMs) \cite{holland1983stochastic,hoff2002latent,snijders2010introduction,wasserman1996logit}. ERGMs are particularly useful for this task because they represent hypotheses about tie formation as explicit model \textit{terms} in an exponential-family model \cite{robins2007introduction,lusher2013exponential}. Fitted parameters show how strongly each included effect is associated with observed ties. ERGMs therefore turn informal explanations of network formation into explicit, testable model specifications.

The remaining difficulty is choosing the \textit{specification}: deciding which structural and attribute effects to include. For instance, even with only 12 available ERGM terms and one term fixed in advance, there are still $2^{11}=2{,}048$ possible specifications. Many combinations are also statistically problematic, leading to unstable estimation or simulated networks that differ sharply from the observed network \cite{snijders2006new,schweinberger2011instability,hunter2008goodness}. Researchers therefore rely on repeated proposal, fitting, diagnosis, and revision \cite{hunter2008ergm,lusher2013exponential}, making ERGM specification a manual trial-and-error process that can require substantial expertise and be difficult to reproduce consistently.

Existing ERGM search and model-selection methods can compare candidate models, but the analyst still has to decide which terms to consider and how to revise a model when it fits poorly. Agentic systems built around large language models (LLMs) are well-suited to this iterative task because they can combine language-based reasoning with external tools and revise their actions using intermediate results \cite{yao2023react,schick2023toolformer}. This closely matches ERGM specification: candidate explanations can be proposed from network context, evaluated through statistical fitting and diagnostics, and revised when the resulting model is inadequate. Recent work also suggests that LLMs can generate and reason about networks shaped by homophily and triadic closure \cite{papachristou2024network,chang2025llms}. Together, these capabilities create an opportunity to use LLMs to guide specification while leaving model fitting and evaluation to statistical tools.

We therefore introduce \textsc{Forge} (Formation-Oriented Reasoning with Guarded ERGMs), an agentic system for ERGM specification. Given a network and a domain description, \textsc{Forge} follows a propose--fit--diagnose--refine cycle. The LLM proposes candidate explanations and model changes, statistical tools fit and evaluate the resulting ERGMs, and fixed rules determine which changes are retained. After refinement, \textsc{Forge} returns the selected ERGM, its diagnostics, and a human-readable interpretation. This reduces manual trial-and-error while keeping model selection tied to statistical evidence.

Our contributions are threefold:
\begin{itemize}

\item We formulate ERGM specification as a constrained search problem that translates hypotheses about tie formation into candidate models while accounting for the network, its attributes, and statistical fit.

\item We introduce \textsc{Forge}, an agentic system that separates model proposal from statistical evaluation. The LLM proposes candidate specifications and local edits, while ERGM tools fit and evaluate the models and fixed rules decide which changes are kept.

\item We evaluate \textsc{Forge} on 16 benchmark networks and 60 controlled networks with known generating specifications. On the benchmark networks, it improves on the best baseline in 13 of 16 cases and matches it in the other three, with a median 24.8\% improvement in goodness of fit. On the controlled networks, it exactly recovers the generating specification in 78.3\% of cases, with 90.8\% term-level F1.

\end{itemize}

\section{Related Work}
\label{sec:related}

\paragraph{ERGM Estimation and Specification.}
Exponential Random Graph Models (ERGMs) represent hypotheses about tie formation as graph statistics in an exponential-family model \cite{wasserman1996logit,robins2007introduction,lusher2013exponential}. Advances in Markov chain Monte Carlo (MCMC) estimation and software such as \texttt{statnet} and \texttt{ergm} have made ERGMs widely used in empirical network analysis \cite{handcock2008statnet,hunter2008ergm,krivitsky2023ergm}. Estimation, however, is only part of the modeling process. Before fitting a model, an analyst must decide which structural and attribute effects to include, and poor combinations can lead to unstable estimation, degeneracy, or poor goodness of fit \cite{hunter2008goodness,schweinberger2011instability}. ERGM analysis therefore often proceeds iteratively: researchers propose a specification, fit it, inspect diagnostics, and revise the included terms \cite{hunter2008ergm,lusher2013exponential}.

A range of methods address parts of this process. Approximate likelihood and improved simulation schemes reduce estimation cost \cite{strauss1990pseudolikelihood,vanDuijn2009pseudolikelihood,hummel2012improving}; stepwise and stochastic search compare candidate models \cite{el2024stochastic}; penalized methods control model complexity \cite{han2024hierarchical}; and Bayesian inference and cross-validation provide alternative criteria for model comparison \cite{caimo2011bayesian,bouranis2018bayesian,yin2019hope}. Most of these methods assume that the candidate terms have already been specified. This leaves an earlier part of the workflow largely manual: deciding which explanations of tie formation are plausible, translating them into ERGM terms, and using failed fits to decide what to try next. \textsc{Forge} focuses on this specification process and uses fitted-model diagnostics to guide subsequent revisions.

\paragraph{Interpretation and Theory Testing with ERGMs.}
The specification problem is especially important because ERGM terms have substantive interpretations. Terms can represent reciprocity, homophily, degree heterogeneity, and other features of tie formation \cite{robins2007introduction,lusher2013exponential}. ERGMs have therefore been used to test theories of friendship \cite{goodreau2007advances,goodreau2009birds}, organizational and political relationships \cite{lazega2001collegial,cranmer2011inferential}, and health-risk networks \cite{morris2008concurrent,goodreau2012bridging}. In these applications, the selected terms determine which effects are tested, while fitted coefficients describe how those effects are associated with observed ties.

This creates a translation problem in both directions. Researchers must first map substantive ideas into appropriate ERGM terms, and after fitting, must translate the selected terms and coefficients back into an interpretable account of the network. Both steps  require domain knowledge and ERGM expertise. \textsc{Forge} makes these steps explicit: the LLM proposes candidate explanations and ERGM terms, statistical tools determine which specifications survive, and the final fitted model is translated into a human-readable interpretation.

\paragraph{LLMs and Structured Generation.}
LLMs provide a natural way to support the first part of this translation. When guided by schemas, tools, or explicit constraints, they can map informal descriptions into structured outputs. Recent work also shows that LLMs can reason about relational settings involving homophily and triadic closure \cite{papachristou2024network,chang2025llms}, while research in text-to-code and semantic parsing maps natural language into executable representations \cite{berant2013semantic,yu2018spider,scholak2021picard}. ERGM specification has an additional requirement: a plausible explanation must map not only to valid syntax, but also to terms that match the network, its attributes, and a model that can be successfully estimated. \textsc{Forge} therefore combines LLM-based generation with deterministic checks and statistical evaluation.

\paragraph{Agentic Systems and Iterative Model Refinement.}
The remaining challenge is that ERGM specification is not a one-step translation task. A candidate model may fail to fit, reproduce the observed network poorly, or suggest that a different combination of terms should be tested. Agentic systems are relevant here because they interleave reasoning with actions and use the resulting observations to guide later decisions \cite{yao2023react,schick2023toolformer}. Related work has explored feedback carried across repeated trials \cite{shinn2023reflexion}, iterative self-revision \cite{madaan2023selfrefine}, and external tools that check intermediate outputs \cite{gou2024critic}. \textsc{Forge} applies this feedback loop to ERGM specification: the LLM proposes a local edit, statistical tools fit and evaluate the edited model, and the resulting diagnostics determine whether the change is kept and what information is available in the next round.

\section{Preliminaries and Problem Formulation}
\label{sec:preliminaries}

We introduce the statistical foundation of Exponential Random Graph Models (ERGMs), then formalize ERGM specification as a constrained search problem. We consider simple, static networks with binary ties and no self-loops. Networks may be directed or undirected, and node attributes $X$ are treated as fixed.

\subsection{ERGM Preliminaries}

Let $G=(V,E)$ denote a network with node set $V$ and edge set $E$, and let $X$ denote fixed node attributes, where each attribute $x\in X$ maps nodes to values, $x:V\rightarrow\mathcal{X}_x$. The attribute space $\mathcal{X}_x$ may be categorical, such as group membership, or numeric, such as age. We represent the network through a random adjacency matrix $Y=(Y_{ij})\in\{0,1\}^{n\times n}$ with $n=|V|$, where $Y_{ij}=1$ indicates an edge from node $i$ to node $j$. We denote the observed realization by $y$.

Given a model specification $M$, an ERGM takes the form
\begin{align}
P_{\theta_M}(Y=y\mid X,M)
&=\exp\left\{
\theta_M^\top s_M(y)
-\psi(\theta_M;X,M)
\right\},
\label{eq:ergm}
\end{align}

where $s_M(y)$ is the vector of network statistics selected by specification $M$, $\theta_M$ is the corresponding parameter vector, and $\psi(\theta_M;X,M)$ is the normalizing constant. Each statistic measures a particular feature of the network. For example, an \texttt{edges} term counts the number of ties, a reciprocity term counts mutually connected pairs in directed networks, and a triadic-closure term measures the extent to which connected nodes share neighbors. Attribute-based terms can similarly measure whether ties occur more often between nodes with the same group membership or similar numeric attributes.

The corresponding coefficient in $\theta_M$ describes how strongly that feature is associated with tie formation, conditional on the other terms in the model. A positive coefficient indicates that networks exhibiting more of that feature receive higher probability, while a negative coefficient indicates the opposite. Thus, the choice of $M$ determines which structural and attribute-based mechanisms the ERGM tests \cite{wasserman1996logit,robins2007introduction,lusher2013exponential}.

ERGM estimation relies on simulated networks because the normalizing constant is generally intractable. Conventional Markov chain Monte Carlo maximum likelihood estimation (MCMLE) approximates the likelihood using MCMC samples, while stochastic approximation (SA) iteratively updates parameter estimates using simulation feedback \cite{snijders2002markov}. A faster but approximate alternative is the maximum pseudolikelihood estimator (MPLE), which replaces the joint likelihood with conditional tie probabilities \cite{strauss1990pseudolikelihood,vanDuijn2009pseudolikelihood}. These estimators provide the fitting and diagnostic quantities used later in our specification procedure.

\subsection{Constrained ERGM Specification}

In most empirical ERGM studies, the statistic vector $s_M(y)$ is selected manually through repeated proposal, fitting, diagnosis, and refinement \cite{hunter2008ergm,lusher2013exponential,goodreau2009birds,schweinberger2011instability}. This becomes difficult when theoretical guidance is limited or many structural and attribute-based terms are plausible. The challenge is not only to choose meaningful mechanisms, but also to find combinations that can be estimated reliably.

Let
\begin{align}
\mathcal{U}(G,X)
&=
\mathcal{U}_{\mathrm{endo}}(G)
\cup
\mathcal{U}_{\mathrm{attr}}(G,X)
\label{eq:term-universe}
\end{align}
denote the valid term universe for an observed network. The structural component $\mathcal{U}_{\mathrm{endo}}(G)$ contains terms available for the network type, while the attribute component $\mathcal{U}_{\mathrm{attr}}(G,X)$ contains terms supported by the available node attributes. The universe determines which terms may be used; a domain description $Q$ provides context for deciding which of these terms are substantively plausible.

A specification $M$ is a subset of the valid term universe
\[
M\subseteq\mathcal{U}(G,X).
\]
Even within this valid term universe, not every subset forms a compatible specification. Some terms may be redundant, incompatible with one another, or excluded by modeling constraints. We therefore define the compatible specification space
\begin{align}
\Omega(G,X)
&=
\left\{
M\subseteq\mathcal{U}(G,X):
\texttt{edges}\in M,\;
C_{\mathrm{comp}}(M;G,X)=1
\right\},
\label{eq:spec-space}
\end{align}
where $C_{\mathrm{comp}}$ checks whether a specification satisfies the required graph, attribute, and term constraints.

If $|\mathcal{U}(G,X)|=p$, requiring an \texttt{edges} term still leaves up to $2^{p-1}$ possible subsets before further restrictions are applied. Exhaustively fitting this space is generally impractical. In addition, a compatible specification may still lead to unstable estimation, degeneracy, or poor goodness of fit \cite{snijders2006new,schweinberger2011instability,hunter2008goodness}.

Let $\mathcal{E}$ denote an ERGM estimation backend. For $M\in\Omega(G,X)$, let $A_{\mathcal E}(M)=1$ indicate that estimation returns finite coefficients, network simulation succeeds, the fitted model passes the prespecified density check, and goodness-of-fit diagnostics can be computed. We call such a specification \emph{eligible}. For an eligible specification, we summarize goodness of fit (GOF) by
\begin{align}
q(M)=\max_{k}|z_k(M)|,
\label{eq:gof}
\end{align}
where $z_k(M)$ is the standardized difference between the observed and simulated value of GOF statistic $k$. Smaller values of $q(M)$ indicate better agreement between the fitted model and the observed network. Using the maximum discrepancy makes the criterion sensitive to the worst reproduced structural feature, so good agreement on several statistics cannot mask a severe mismatch on another.

Conceptually, the specification problem can be written as
\begin{align}
M^\star
\in
\arg\min_{\substack{M\in\Omega(G,X)\\A_{\mathcal E}(M)=1}}
q(M).
\label{eq:constrained-search}
\end{align}
Searching this space exactly would require fitting a large number of candidate models. \textsc{Forge} instead searches for a low-discrepancy eligible specification without exhaustively enumerating $\Omega(G,X)$.

\paragraph{Problem statement.}
\textit{Given an observed network $G=(V,E)$, node attributes $X$, a domain description $Q$, and valid term universe $\mathcal{U}(G,X)$, identify a substantively plausible ERGM specification that is eligible under $\mathcal{E}$ and achieves low GOF discrepancy $q(M)$ without exhaustively enumerating $\Omega(G,X)$.}

This formulation separates two parts of the task: deciding which mechanisms are worth considering, and testing whether the resulting specification works statistically. We next describe how \textsc{Forge} combines these two steps.

\section{Approach}
\label{sec:approach}

\begin{table*}[t]
\centering
\caption{LLM roles in \textsc{Forge}: Stage~I generates candidate specifications, Stage~III proposes one local edit from statistical feedback, and Stage~IV interprets the fixed fitted model. Stage~II contains no LLM call; statistical software handles statistical screening and selection. Prompts (\textit{LLM task}) are shortened for readability.}
\label{tab:llm-interfaces}
\small
\setlength{\tabcolsep}{5pt}
\begin{tabularx}{\textwidth}{@{}lXXX@{}}
\toprule
\textbf{Stage} & \textbf{Information given to the LLM} & \textbf{LLM task} & \textbf{Returned output} \\
\midrule

Stage I &
Domain description, network summary and diagnostics, available node attributes, and the valid term universe &
\textit{Identify plausible formation mechanisms from the observed evidence, map each mechanism to a term in the valid term universe, and assemble one to three complete candidate specifications.} &
For each proposal: formation mechanism, supporting evidence, mapped ERGM term, and 1--3 ranked candidate specifications \\

\addlinespace

Stage III &
Current specification, current $q(M)$, fit diagnostics, largest GOF residuals, previously rejected edits, and the valid term universe &
\textit{Use the statistical feedback to propose exactly one local edit. The edit must add, remove, or replace one ERGM term. Do not decide whether the edited model should be accepted.} &
Exactly one \texttt{add}, \texttt{remove}, or \texttt{replace} action, together with a short reason tied to the supplied diagnostics \\

\addlinespace

Stage IV &
Final selected specification, fitted coefficients, and coefficient directions &
\textit{Map each fitted ERGM term to its formation mechanism and explain what the fitted direction means for the observed network. Do not make causal claims.} &
For each fitted term: mechanism label, coefficient direction, human-readable interpretation, and limitation \\

\bottomrule
\end{tabularx}
\end{table*}

\subsection{System Overview}

Figure~\ref{fig:forge-framework} shows the overall \textsc{Forge} workflow, which has four stages. Stage~I uses the LLM to reason about plausible explanations for how ties form and translate them into a small set of candidate ERGM specifications. Stage~II fits and evaluates these candidates using statistical tools, removes models that fail the required checks, and selects the best candidate that passes them. Stage~III iteratively refines the selected model through an agentic feedback loop. At each refinement round, the LLM receives the current statistical results and proposes one change to the specification. The revised model is then fitted and evaluated, and the change is kept if it passes the required checks and improves goodness of fit. Stage~IV takes the final fitted model and translates it into a human-readable interpretation of how ties form in the observed network.

In Stages~I and IV, the LLM is used for initial candidate generation and final interpretation, respectively, while Stage~II contains no LLM call. Stage~III forms an agentic feedback loop, where statistical results from one attempt inform the LLM's next proposed modification. Model fitting, statistical checks, and acceptance decisions remain outside the LLM throughout the workflow. Because fitting and simulation are handled through a separate ERGM backend, the backend can be replaced without changing the overall search procedure. In our main experiments, we use the stochastic-approximation estimator implemented in Statnet's \texttt{ergm} package.

Next, we describe each stage and the role of the LLM in more detail. Algorithm~\ref{alg:forge} gives the Stage~II--III selection and refinement procedure.

\subsection{\textsc{Forge} Workflow}

\paragraph{Stage I: Candidate Specification Generation.}
Stage~I turns the network context into a small set of candidate ERGM specifications. The LLM first identifies plausible formation mechanisms from the supplied evidence, maps each mechanism to an exact term in $\mathcal U(G,X)$, and then assembles up to three ranked specifications. Limiting the output to three gives Stage~II several plausible alternatives to compare without generating a large candidate pool. For example, high clustering may suggest triadic closure and therefore \texttt{gwesp}, while frequent within-group ties may suggest homophily and therefore \texttt{nodematch}. This mechanism-first step ties each proposed ERGM term to a clear hypothesis about tie formation.

\paragraph{Stage II: Statistical Screening and Selection.}
Stage~II first removes specifications that are incompatible with the network or are known to be unnecessarily fragile. Every candidate includes \texttt{edges}, which represents the baseline propensity for a tie and allows structural and attribute terms to be evaluated relative to a common density effect~\citep{robins2007introduction,lusher2013exponential}.
All remaining terms must belong to the valid term universe $\mathcal U(G,X)$ and match the observed network. Directed terms such as \texttt{mutual} are therefore unavailable for undirected networks, while attribute terms are included only when the required categorical or numeric covariate is present. We also restrict each geometrically weighted term family to a single decay value to avoid including redundant versions of the same structural effect.

\textsc{Forge} excludes the raw \texttt{triangle} term from its search space. Triangle counts introduce strong dependence because one added edge can close several triangles at once, and such specifications are well known to be vulnerable to ERGM degeneracy and concentration on unrealistic near-empty or near-complete graphs~\citep{snijders2006new,schweinberger2011instability}. Triadic closure remains available through geometrically weighted shared-partner terms such as \texttt{gwesp}, which give diminishing weight to additional shared partners and are commonly used to model closure more stably~\citep{snijders2006new,robins2007introduction,lusher2013exponential}. These restrictions are fixed before LLM generation, and incompatible outputs are rejected.

Candidates that pass this initial screen are fitted through $\mathcal E$. Each candidate first receives a density check that compares the density of simulated networks with the observed network, providing an inexpensive screen for unstable fits before the full GOF evaluation. Candidates that pass this check are then evaluated using the same simulation-based GOF procedure. We retain MPLE-based pseudo-BIC (PBIC) as a secondary diagnostic; it applies the Bayesian information criterion (BIC) to the pseudolikelihood and penalizes additional terms. Candidates that pass all post-fit checks form the eligible set $\mathcal C_{\mathrm{eligible}}$.
The eligible candidate with the lowest GOF discrepancy is then passed to Stage~III:
\begin{equation}
M_0=\arg\min_{M\in\mathcal C_{\mathrm{eligible}}}q(M).
\label{eq:forge-start}
\end{equation}
The model with the lowest $q(M)$ has the smallest worst-case structural mismatch among the eligible candidates.

\paragraph{Stage III: Agentic Specification Refinement.}
Stage~III starts from the model selected in Stage~II and refines it through a sequence of local edits and statistical feedback. At refinement round $t$, the current state is
\begin{equation}
S_t=(M_t,D_t,H_t),
\end{equation}
where $M_t$ is the current specification, $D_t$ summarizes its fit diagnostics, and $H_t$ records earlier edits and their outcomes.

Given this state and the original network context, the LLM proposes exactly one local edit $a_t$: adding, removing, or replacing a single ERGM term. Restricting each round to one edit makes it clear which specification change is responsible for any change in model fit. The LLM receives the current $q(M_t)$, the largest GOF mismatches, the valid term universe, and the history of rejected edits. It receives this compact diagnostic summary rather than raw estimator output, and previously rejected edits are excluded so that later rounds do not repeat changes that have already failed the statistical checks.

Applying $a_t$ produces a revised specification $M'$. The same compatibility and post-fit checks used in Stage~II are then applied to $M'$, which is fitted and evaluated through backend $\mathcal E$. The edit is retained only if the revised model is eligible and has a lower GOF discrepancy than the current model. Otherwise, $M_t$ is kept and the rejected edit is added to $H_t$, providing feedback for the next round.

We run at most four refinement rounds, limiting the number of additional ERGM fits and keeping the refinement budget fixed across networks. Let $M_T$ denote the final retained specification. Because Stage~III keeps only edits that reduce $q(M)$,
\begin{equation}
q(M_T)\leq q(M_0)
=\min_{M\in\mathcal C_{\mathrm{eligible}}}q(M).
\label{eq:nondegradation}
\end{equation}
Stage~III can therefore improve the model selected in Stage~II or leave it unchanged. If no proposed edit improves the current model, the Stage~II selection remains the final specification.

Algorithm~\ref{alg:forge} summarizes the full search procedure. Here, $\operatorname{Eval}_{\mathcal E}$ fits and evaluates a specification, $\operatorname{LLMEdit}$ proposes one local edit, and $\operatorname{Edit}$ applies it.

\begin{algorithm}[t]
\caption{\textsc{Forge} Agentic Specification Search}
\label{alg:forge}
\small
\begin{algorithmic}[1]
\REQUIRE Candidate specifications $\mathcal C$; backend $\mathcal E$; network $(G,X)$; context $Q$; maximum rounds $T$
\ENSURE Refined specification $M_T$ and search history $H_T$
\STATE $\mathcal C_{\mathrm{eligible}}\leftarrow\varnothing$
\FOR{$M\in\mathcal C$}
    \IF{$M\in\Omega(G,X)$}
        \STATE $D_M\leftarrow\operatorname{Eval}_{\mathcal E}(M;G,X)$
        \IF{$A_{\mathcal E}(M)=1$}
            \STATE $\mathcal C_{\mathrm{eligible}}\leftarrow\mathcal C_{\mathrm{eligible}}\cup\{M\}$
        \ENDIF
    \ENDIF
\ENDFOR
\STATE $M_0\leftarrow\arg\min_{M\in\mathcal C_{\mathrm{eligible}}}q(M)$
\STATE $D_0\leftarrow D_{M_0}$; $H_0\leftarrow\varnothing$
\FOR{$t=0,\ldots,T-1$}
    \STATE $S_t\leftarrow(M_t,D_t,H_t)$
    \STATE $a_t\leftarrow\operatorname{LLMEdit}(Q,G,X,S_t)$ \COMMENT{add, remove, or replace}
    \STATE $M'\leftarrow\operatorname{Edit}(M_t,a_t)$
    \IF{$M'\in\Omega(G,X)$}
        \STATE $D'\leftarrow\operatorname{Eval}_{\mathcal E}(M';G,X)$
        \IF{$A_{\mathcal E}(M')=1$ \AND $q(M')<q(M_t)$}
            \STATE $M_{t+1}\leftarrow M'$; $D_{t+1}\leftarrow D'$
        \ELSE
            \STATE $M_{t+1}\leftarrow M_t$; $D_{t+1}\leftarrow D_t$
        \ENDIF
    \ELSE
        \STATE $D'\leftarrow\text{incompatible}$
        \STATE $M_{t+1}\leftarrow M_t$; $D_{t+1}\leftarrow D_t$
    \ENDIF
    \STATE $H_{t+1}\leftarrow H_t\cup\{(a_t,M',D',M_{t+1})\}$
\ENDFOR
\RETURN $M_T,H_T$
\end{algorithmic}
\end{algorithm}

\paragraph{Stage IV: Post-fit Interpretation.}
Stage~IV runs only after model selection is complete. Given the final specification and fitted coefficients, the LLM maps each selected term to its formation mechanism and gives a human-readable interpretation of the estimated effect. For example, \texttt{mutual} is described as a tendency for ties to be reciprocated, \texttt{gwesp} as a tendency for ties to form between nodes with shared neighbors, and \texttt{nodematch} as a tendency for ties to occur between nodes with the same attribute value. This final step turns ERGM-specific terminology and coefficient signs into a concise interpretation that can be read without detailed knowledge of ERGM term syntax.

Stage~IV is purely interpretive: it does not modify the selected specification, fitted coefficients, or GOF results. Section~\ref{sec:llm-choice} evaluates whether these interpretations remain consistent across five language models.

\subsection{LLM Inputs and Outputs}

Table~\ref{tab:llm-interfaces} summarizes the inputs, instructions, and structured outputs for the three LLM-facing stages; prompt templates and output schemas are provided in the supplement.

\section{Experiments}

\subsection{Experimental Settings}
\label{sec:experimental-settings}

\paragraph{Benchmarks.}
We evaluate \textsc{Forge} on 16 benchmark networks spanning friendship, communication, organizational, school, trust, and online interaction networks \cite{traud2012social,snapEmailEU,ergm490,rsiena150,networkDynamicData030,concorR021,lazega2001collegial,noordinTopData,snapCollegeMsg,snapBitcoinAlpha,snapTwitchRU}. The networks range from 16 to 4,385 nodes and from 20 to 37,304 edges, with valid term universes containing 8--43 terms. After fixing the required \texttt{edges} term, the search spaces range from 128 to $4.40\times10^{12}$ raw specifications. Table~\ref{tab:benchmarks} lists all 16 benchmark networks. Full source and preprocessing details are provided in the supplement.

ERGM estimation becomes increasingly expensive as networks grow because likelihood-based fitting relies on repeated network simulation \cite{snijders2002markov,hummel2012improving}. We therefore use MCMLE as the conventional likelihood-based reference on the 12 networks where repeated fitting remains practical, and stochastic approximation (SA) as the common estimator across all 16 benchmark networks. To separate estimator behavior from specification quality, we compare the estimators on fixed specifications. We additionally use Wiki-Vote, a directed network of Wikipedia users voting on administrator promotions, as a larger-scale stress test. Wiki-Vote (7,115 nodes, 103,689 edges) is substantially larger than the 16 benchmarks \cite{snapWikiVote}. We compare SA with equilibrium expectation (EE), an estimator developed for faster ERGM fitting on large networks \cite{byshkin2018fast}. Wiki-Vote is not included in the 16-network effectiveness comparison. The \textsc{Forge} search procedure remains unchanged across estimators. A complementary 12-network MCMLE/BIC robustness summary is reported in the supplement.

\begin{table}[t]
\centering
\caption{16 benchmark networks used for evaluating FORGE, spanning 16--4,385 nodes and 20--37,304 edges, with valid term universes containing 8--43 terms.
MCMLE/SA indicates that both estimators are evaluated. Wiki-Vote is
evaluated separately as a scalability stress test.}
\label{tab:benchmarks}
\renewcommand{\arraystretch}{0.92}
\setlength{\tabcolsep}{2.5pt}
\resizebox{\columnwidth}{!}{%
\begin{tabular}{lrcrrrl}
\toprule
Network & $|V|/|E|$ & Dir. & Attr. & $|\mathcal{U}|$ & Raw specs. & Est. \\
\midrule
\multicolumn{7}{l}{\textit{Smaller benchmark networks}} \\
Florentine   & 16/20     & No  & 3  & 14 & 8,192                 & MCMLE/SA \\
Krackhardt   & 21/190    & Yes & 4  & 23 & $4.19{\times}10^6$    & MCMLE/SA \\
Kapferer     & 39/158    & No  & 0  & 8  & 128                   & MCMLE/SA \\
Glasgow s50  & 50/113    & Yes & 4  & 25 & $1.68{\times}10^7$    & MCMLE/SA \\
Lazega       & 71/1,574  & Yes & 8  & 43 & $4.40{\times}10^{12}$ & MCMLE/SA \\
Noordin Top  & 79/200    & No  & 15 & 40 & $5.50{\times}10^{11}$ & MCMLE/SA \\
Manufacturing& 167/5,783 & Yes & 0  & 12 & 2,048                 & MCMLE/SA \\
Enron        & 184/3,010 & Yes & 1  & 15 & 16,384                & MCMLE/SA \\
Faux Mesa    & 205/203   & No  & 3  & 16 & 32,768                & MCMLE/SA \\
Faux Dixon   & 248/978   & No  & 3  & 17 & 65,536                & MCMLE/SA \\
\addlinespace[2pt]
\multicolumn{7}{l}{\textit{Larger benchmark networks}} \\
Caltech36     & 769/16,656    & No  & 7 & 20 & 524,288 & MCMLE/SA \\
Email-Eu-core & 1,005/24,929  & Yes & 1 & 13 & 4,096   & SA \\
Faux Magnolia & 1,461/974     & No  & 3 & 16 & 32,768  & MCMLE/SA \\
CollegeMsg    & 1,899/20,296  & Yes & 1 & 16 & 32,768  & SA \\
Bitcoin-Alpha & 3,783/24,186  & Yes & 0 & 12 & 2,048   & SA \\
Twitch-RU     & 4,385/37,304  & No  & 4 & 18 & 131,072 & SA \\
\addlinespace[2pt]
\multicolumn{7}{l}{\textit{Separate scalability stress test}} \\
Wiki-Vote     & 7,115/103,689 & Yes & 0 & 12 & 2,048   & SA/EE \\
\bottomrule
\end{tabular}%
}
\end{table}

\begin{table*}[t]
\centering
\caption{Full results on the 16 benchmark networks. Lower $q(M)$ indicates better fit. \textsc{Forge} before refinement is the model selected after Stage~II, and \textsc{Forge} Final is the model after Stage~III. Overall gain is measured against the best non-\textsc{Forge} baseline that passes the statistical checks. Negative $\Delta$PBIC indicates improvement. Because $q(M)$ is network-specific, cross-network comparisons use relative reductions.}
\label{tab:all-network-results}

\footnotesize
\setlength{\tabcolsep}{2.8pt}
\renewcommand{\arraystretch}{0.97}

\begin{tabular*}{\textwidth}{@{\extracolsep{\fill}}llrrrrrrr@{}}
\toprule
& & \multicolumn{3}{c}{\textbf{GOF discrepancy $q(M)$}} &
\multicolumn{1}{c}{\textbf{Improvement}} &
\multicolumn{3}{c}{\textbf{Complexity and cost}} \\
\cmidrule(lr){3-5}\cmidrule(lr){6-6}\cmidrule(lr){7-9}
Network &
Type &
\makecell{\textbf{Best}\\\textbf{baseline}} &
\makecell{\textbf{\textsc{Forge}}\\\textbf{before refinement}} &
\makecell{\textbf{\textsc{Forge}}\\\textbf{Final}} &
\makecell{\textbf{Overall}\\\textbf{gain}} &
\makecell{\textbf{Terms}\\Baseline$\to$Final} &
\makecell{\textbf{$\Delta$PBIC}\\Final$-$Baseline} &
\makecell{\textbf{Stage III}\\\textbf{time (s)}} \\
\midrule

\multicolumn{9}{l}{\textit{Smaller benchmark networks}} \\
\addlinespace[1pt]
Faux Mesa      & School       & 3.56 (FS)   & 2.92 (F)    & \textbf{2.88}   & \textbf{19.2\%} & 4$\to$6 & +7.78          & 32 \\
Faux Dixon     & School       & 3.93 (OS)   & 3.57 (F)    & \textbf{3.57}   & \textbf{9.1\%}  & 4$\to$4 & $-74.31$       & 53 \\
Kapferer       & Organization & 6.58 (UC)   & 6.58 (UC)   & \textbf{6.58}   & 0.0\%            & 3$\to$3 & 0.00           & 39 \\
Lazega         & Organization & 59.12 (UC)  & 42.02 (F)   & \textbf{42.02}  & \textbf{28.9\%} & 5$\to$6 & $-242.64$      & 107 \\
Krackhardt     & Organization & 16.16 (EO)  & 16.16 (EO)  & \textbf{16.16}  & 0.0\%            & 1$\to$1 & 0.00           & 30 \\
Glasgow        & Friendship   & 13.31 (RK)  & 2.69 (F)    & \textbf{1.68}   & \textbf{87.4\%} & 4$\to$8 & $-500.99$      & 29 \\
Manufacturing  & Email        & 945.10 (EO) & 945.10 (EO) & \textbf{291.25} & \textbf{69.2\%} & 1$\to$2 & $-17{,}732.17$ & 130 \\
Enron          & Email        & 374.18 (RK) & 73.92 (F)   & \textbf{29.41}  & \textbf{92.1\%} & 4$\to$6 & $-2{,}014.51$  & 69 \\
Florentine     & Family       & 1.34 (RK)   & 1.34 (RK)   & \textbf{1.32}   & \textbf{1.7\%}  & 4$\to$4 & +0.04          & 31 \\
Noordin        & Covert       & 3.91 (OS)   & 2.48 (F)    & \textbf{2.48}   & \textbf{36.6\%} & 4$\to$6 & $-41.82$       & 47 \\

\addlinespace[5pt]
\multicolumn{9}{l}{\textit{Larger benchmark networks}} \\
\addlinespace[1pt]
Caltech36      & School    & 4969.90 (OS) & 65.80 (F)  & \textbf{65.80} & \textbf{98.7\%} & 4$\to$7 & $-4{,}006.42$ & 2,255 \\
Email-Eu-core  & Email     & 40.02 (FS)   & 40.02 (FS) & \textbf{36.42} & \textbf{9.0\%}  & 4$\to$5 & $-2{,}031.20$ & 489 \\
Faux Magnolia  & School    & 127.97 (EO)  & 5.54 (F)   & \textbf{5.54}  & \textbf{95.7\%} & 1$\to$5 & $-4{,}618.07$ & 51 \\
CollegeMsg     & Messaging & 47.63 (FS)   & 33.13 (F)  & \textbf{33.13} & \textbf{30.4\%} & 4$\to$6 & $-1{,}477.18$ & 832 \\
Bitcoin-Alpha  & Trust     & 35.68 (FS)   & 35.68 (FS) & \textbf{35.68} & 0.0\%            & 4$\to$4 & 0.00          & 618 \\
Twitch-RU      & Streaming & 29.20 (UC)   & 29.20 (UC) & \textbf{23.16} & \textbf{20.7\%} & 5$\to$6 & $-813.90$     & 1,955 \\

\bottomrule
\end{tabular*}

\vspace{3pt}
\begin{minipage}{0.99\textwidth}
\scriptsize
\textit{Notes.} EO = edges-only, RK = Random-$K$, OS = one-shot, FS = few-shot, UC = unconstrained, and F = a candidate generated by \textsc{Forge}. Refinement time includes proposal, fitting, and GOF evaluation.
\end{minipage}
\end{table*}

\paragraph{Baselines.}
We compare \textsc{Forge} with five baselines: \textit{edges-only}, Random-$K$, and three direct LLM baselines. The \textit{edges-only} model provides the simplest reference, while Random-$K$ adds three randomly selected non-edge terms using a fixed seed. \textit{One-shot} and \textit{few-shot} each produce a four-term ERGM, with few-shot additionally receiving examples of evidence-to-term mappings. The \textit{unconstrained} baseline allows the LLM to choose between 3-8 terms.

To compare the initial LLM proposals fairly, both \textsc{Forge} and direct prompting use three candidates. \textsc{Forge} generates three Stage~I candidates, while direct prompting contributes one candidate each from one-shot, few-shot, and unconstrained prompting. This gives both approaches the same number of initial attempts. The edges-only and Random-$K$ baselines are included in the full Stage~II comparison. All LLM-based methods receive the same network information and choose from the same valid term universe.

In the full \textsc{Forge} workflow, Stage~II considers the three \textsc{Forge} candidates together with the baseline models and selects the model with the lowest $q(M)$ that passes the statistical checks. Stage~III then tries up to four one-term changes and keeps a change only when it further reduces $q(M)$. Because the baseline models are included in Stage~II, \textsc{Forge} can keep the best baseline when it does not find a better model. We therefore focus on how often the full workflow improves beyond that baseline.

\paragraph{Common controls.}
Beyond using the same LLM, we also keep the network-specific information available to each method fixed. All LLM-based methods receive the dataset and domain description, numbers of nodes and edges, directionality, density, reciprocity when applicable, global clustering, degree variation, available node attributes, and the valid term universe.

We use the 60 controlled networks to choose the LLM used in the 16-network benchmark. This choice is based on Stage~I candidate generation. Among five tested LLMs, GPT-5.4 \cite{openai2026gpt54} and Llama 4 Maverick \cite{meta2025llama4} most often rank the generating specification first. GPT-5.4 has the higher term-level F1 and produces compatible specifications on all 60 networks. We then use GPT-5.4 for one-shot, few-shot, unconstrained, and \textsc{Forge} throughout the 16-network benchmark, so differences between these methods come from the specification procedure rather than from using different LLMs. Additional LLM results are reported in the supplement.

\paragraph{Evaluation protocol.}
After fitting an ERGM, we ask whether networks simulated from it resemble the observed network. We measure this using the simulation-based goodness-of-fit score $q(M)$ from Eq.~\eqref{eq:gof}. For each network statistic, we compare the observed value with the values obtained from simulated networks and scale the difference by its simulated variation. We examine degree distributions, in- and out-degree distributions for directed networks, shared-partner counts, and geodesic distances. The score $q(M)$ records the largest absolute difference across these statistics, so lower values indicate a better fit. We use the maximum because an ERGM that fits most statistics well can still reproduce one important feature of the observed network very poorly. Because these statistics differ across networks, we use $q(M)$ to compare models within the same network. Across networks, we compare relative improvements instead of raw $q(M)$ values.

For the 16-network benchmark, each candidate first receives a 30-simulation density check to screen out clearly poor fits. Candidates that pass are then evaluated using 100 simulated networks, with the same simulation seed used for competing models within each network. We use $q(M)$ to select the starting model for Stage~III and to decide whether each Stage~III edit is kept. We also report the root-mean-square GOF discrepancy, the 95th percentile of absolute GOF discrepancies, and PBIC as supporting measures. Stage~III runs for at most four rounds, with each round adding, removing, or replacing one term. For each network, overall gain is defined as
$100[q(M_{\mathrm{baseline}})-q(M_{\mathrm{final}})]/q(M_{\mathrm{baseline}})$.

\paragraph{Controlled recovery benchmark.}
Because the 16 benchmark networks do not have known generating specifications, we use a separate controlled benchmark with 60 networks whose generating specifications are known. We consider six settings representing reciprocity, homophily, triadic closure, degree heterogeneity, homophily with closure, and reciprocity with homophily and closure. For each setting, we generate ten independent 120-node networks, giving 60 networks in total.

For each network, the LLM receives the same type of network summary and node-attribute information used in the 16-network benchmark, but is not told which ERGM terms or parameter values generated the network. It must therefore recover the generating specification from the observed network information alone.

All methods choose from the same fixed term set. It contains \texttt{edges}, \texttt{mutual} for directed networks, and \texttt{nodematch("group")}.\\It also includes \texttt{gwesp(0.5)} and \texttt{gwdegree(0.5)}. Any compatible subset may be selected, so exact recovery requires choosing the correct terms without adding unnecessary ones. For \textsc{Forge}, recovery is evaluated on the final specification after Stage~III, while the direct prompting baselines are evaluated on their generated specifications. A case is an exact recovery when the evaluated specification matches the generating specification. We also report term-level precision, recall, and F1, together with accuracy in identifying whether each fitted effect has the correct positive or negative direction.

\paragraph{Post-fit interpretation benchmark.}
Finally, we test whether different LLMs interpret the same fitted ERGM consistently. We give five LLMs the same six fitted models, including their selected terms and coefficient directions. Because the models are already fixed, any differences come from interpretation rather than model selection or fitting. We evaluate whether each LLM correctly explains what the selected terms represent, identifies whether each effect is positive or negative, and refers to the correct ERGM terms.

\subsection{Experimental Results}
\label{sec:experimental-results}

\subsubsection{\textsc{Forge} improves on the best baseline on 13 of 16 networks.}
\label{sec:main-effectiveness}

Table~\ref{tab:all-network-results} reports the full comparison across all 16 benchmark networks. The full \textsc{Forge} workflow improves on the best baseline on 13 networks and matches it on the remaining three, with a median 24.8\% reduction in $q(M)$. The gains appear across both parts of the benchmark: 8 of 10 smaller networks and 5 of 6 larger networks improve. The largest reductions occur on Caltech36 (98.7\%), Faux Magnolia (95.7\%), and Enron (92.1\%), showing that \textsc{Forge} can substantially improve models with large initial GOF discrepancies.

\textsc{Forge} also tends to select richer specifications, with the median number of terms increasing from 4.0 to 5.5. PBIC, which penalizes additional terms, nevertheless improves on 11 networks, ties on three, and worsens on only two. Thus, the selected models generally improve both structural fit and a complexity-adjusted criterion.

The table also shows how these gains develop through the workflow. After Stage~II, a \textsc{Forge}-generated candidate already outperforms the best baseline on 9 of 16 networks; after Stage~III, this increases to 13 of 16. We next examine how statistical feedback produces these additional improvements.

\subsubsection{\textsc{Forge} produces stronger initial candidates and improves them through refinement.}

Before refinement, \textsc{Forge} already produces stronger candidates than direct prompting. Each approach has three candidates: \textsc{Forge} generates three Stage~I candidates, while direct prompting uses one-shot, few-shot, and unconstrained prompting. Among the 11 networks where both produce at least one candidate that passes the statistical checks, the best \textsc{Forge} candidate performs better on seven networks and the best direct candidate on four. Stage~II then evaluates these candidates together with the other baselines and selects the best-fitting model for Stage~III.

Stage~III uses statistical feedback to decide which proposed changes are worth keeping. It accepts seven edits, improving seven different models with a median 20.7\% reduction in $q(M)$. This shows that the statistical checks retain useful revisions rather than accepting LLM suggestions automatically.

\begin{figure}[!t]
\centering
\includegraphics[width=\columnwidth]{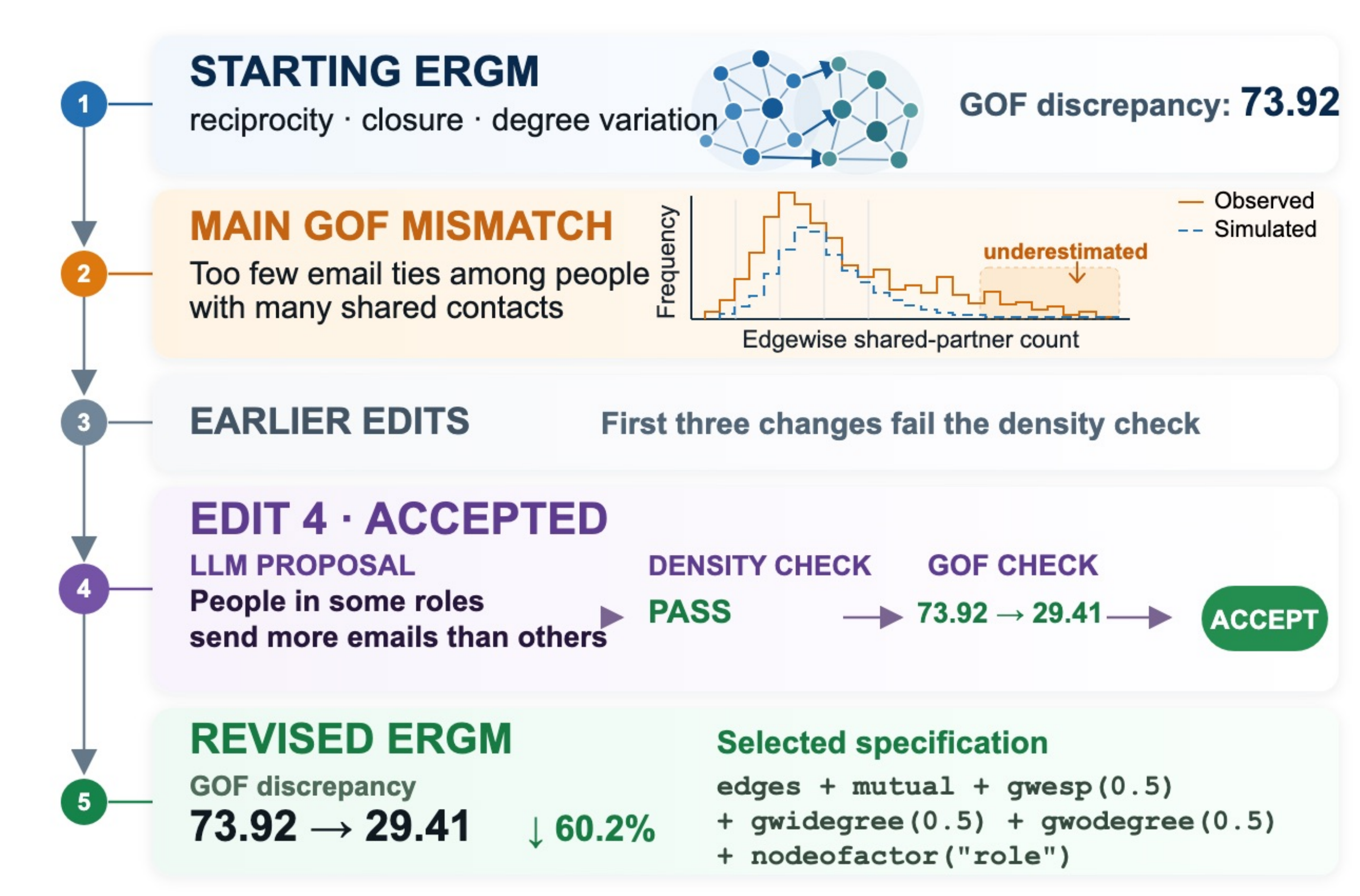}
\caption{Stage~III refinement on Enron. The first three proposed edits fail the density check and are rejected. The fourth accounts for differences in email activity across roles, passes the statistical checks, and reduces $q(M)$ from 73.92 to 29.41 (60.2\%). Other GOF measures improve at the same time.}
\Description{Enron refinement trajectory across four Stage~III edits. The first three edits fail the density check and leave $q(M)$ at 73.92; the fourth is accepted and lowers $q(M)$ to 29.41, a 60.2\% reduction, while the other diagnostics also improve.}
\label{fig:enron-trajectory}
\end{figure}

Figure~\ref{fig:enron-trajectory} shows this process on the Enron email network. The first three proposed edits fail the density check and leave the model unchanged. The fourth accounts for differences in email activity across employee roles and passes the statistical checks, reducing $q(M)$ from 73.92 to 29.41, a 60.2\% improvement. The other GOF measures and PBIC improve as well.

\subsubsection{\textsc{Forge} recovers the generating specification in 78.3\% of controlled networks.}
\label{sec:controlled-recovery}

\begin{figure}[!t]
\centering
\includegraphics[width=0.85\columnwidth]{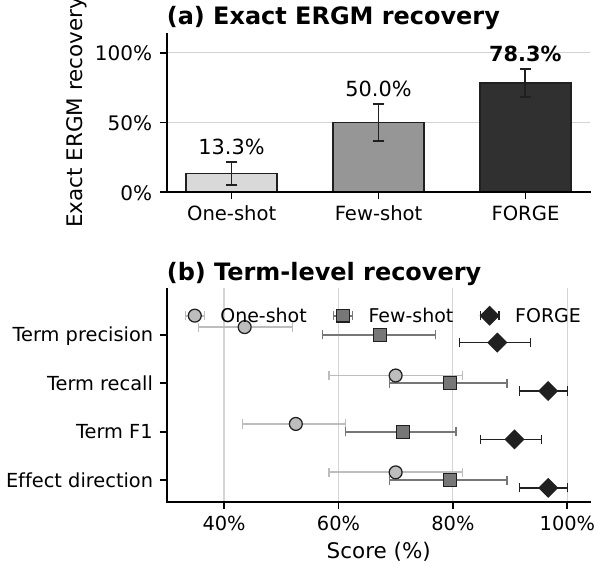}
\caption{Recovery on 60 controlled networks with known generating specifications. (a) Exact specification recovery for \textsc{Forge}, few-shot, and one-shot prompting. (b) Term-level precision, recall, F1, and effect-direction accuracy. Error bars show 95\% bootstrap confidence intervals.}
\Description{The upper panel compares exact ERGM specification recovery for one-shot, few-shot, and \textsc{Forge}. The lower panel compares term-level precision, recall, F1, and effect-direction accuracy across the three methods.}
\label{fig:controlled-recovery}
\end{figure}

On the 16 benchmark networks, we can measure how well an ERGM fits, but the generating specification is unknown. We therefore evaluate \textsc{Forge} on 60 controlled networks whose generating specifications are known but hidden from the LLM.
\textsc{Forge} exactly recovers the generating specification in 78.3\% of cases, compared with 50.0\% for few-shot prompting and 13.3\% for one-shot prompting. Even when the full specification is not recovered exactly, the selected terms remain highly accurate: \textsc{Forge} reaches 90.8\% term-level F1 and identifies the correct positive or negative effect direction in 96.7\% of cases. Figure~\ref{fig:controlled-recovery} summarizes both the exact and term-level results.

\subsubsection{LLMs differ most on exact specifications and term references.}
\label{sec:llm-choice}

\begin{figure}[!t]
\centering
\includegraphics[width=0.9\columnwidth]{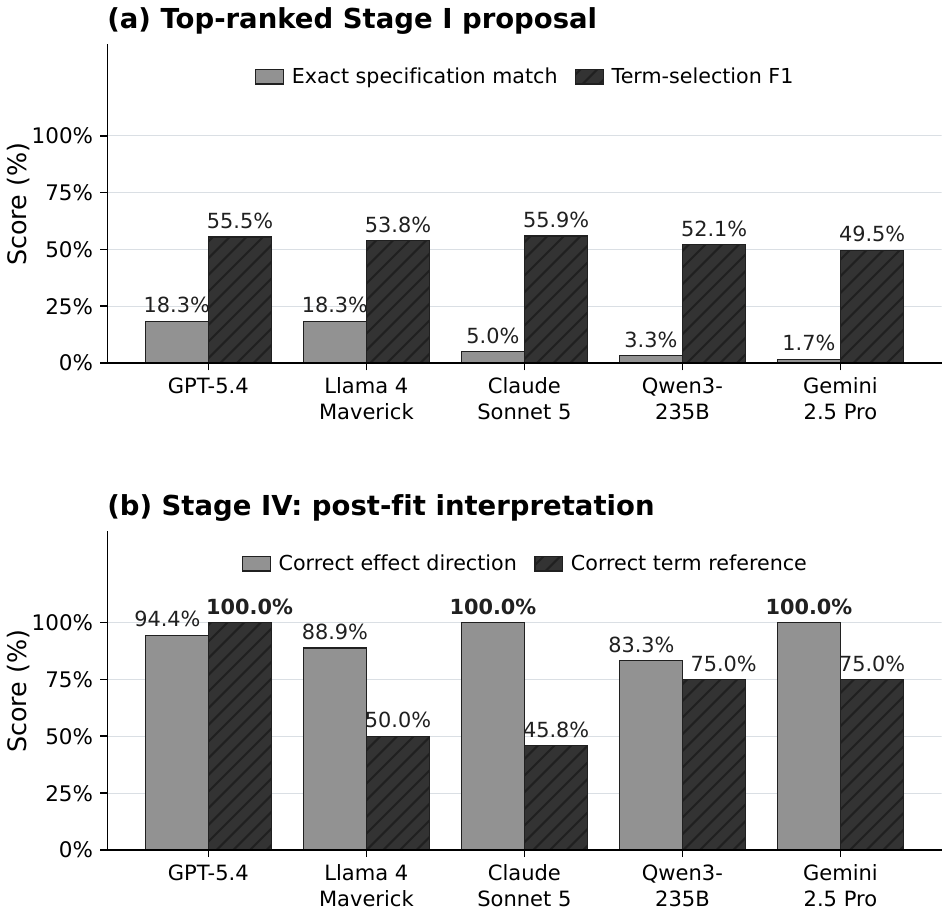}
\caption{Comparison of five LLMs at two points in \textsc{Forge}. (a) The Stage~I candidate ranked first by each LLM before statistical screening and refinement, evaluated by exact specification match and term-level F1. (b) Interpretation of the same six fitted ERGMs, evaluated by effect-direction and term-reference accuracy. Higher values are better.}
\Description{Two bar charts compare five LLMs. The upper panel reports exact specification recovery and term-level F1 for the first-ranked Stage~I proposal. The lower panel reports effect-direction and term-reference accuracy for six fixed fitted ERGMs.}
\label{fig:llm-comparison}
\end{figure}

We compare five LLMs at two points in \textsc{Forge}. During candidate generation, the five LLMs achieve similar term-level F1 but differ more in how often they rank the generating specification first. GPT-5.4 and Llama 4 Maverick perform best, each doing so on 11 of 60 controlled networks. GPT-5.4 achieves the higher term-level F1, so we use it in the main experiments.
After fitting, all five LLMs usually identify whether an effect is positive or negative, with accuracy ranging from 83.3\% to 100\%. They differ more in correctly referring to the ERGM terms provided to them, with accuracy ranging from 45.8\% to 100\%. Figure~\ref{fig:llm-comparison} summarizes both comparisons.

\subsubsection{Faster estimators give similar results at lower cost.}
\label{sec:estimator-results}

We first compare SA with conventional MCMLE on the same ERGM specifications to see whether the choice of estimator changes the fitted model. Where both complete, their coefficient estimates differ by less than 0.3\% and have the same signs. SA is also substantially faster: on Email-Eu-core, it finishes in 5.05 seconds, while MCMLE does not finish within one hour.

The main results are also similar when we use conventional MCMLE and BIC. On the 12-network MCMLE/BIC evaluation, \textsc{Forge} achieves the best or tied BIC on 8 of the 11 completed networks and selects the same ERGM as the best baseline on one more; additional details are provided in the supplement.

Faster estimators can also make larger networks more practical. On Wiki-Vote, with 7,115 nodes and 103,689 edges, EE reduces fitting time from 67.31 seconds with SA to 18.15 seconds, a 3.7$\times$ speedup. The fitted model still has poor goodness of fit, showing that faster estimation reduces computational cost but does not replace the need for a good specification.

\section{Discussion}
\label{sec:discussion}

A central design choice in \textsc{Forge} is the separation between proposing a specification and deciding whether to keep it. The LLM proposes candidate explanations and local edits, while compatibility checks, ERGM fitting, simulation, and GOF determine which changes are retained. Because many proposed edits are rejected, statistical feedback guides the search rather than serving only as a final validation step.

The two benchmarks answer different questions. On the 16 benchmark networks, the generating specification is unknown, so we ask whether \textsc{Forge} finds models that reproduce the observed network better than the baselines. On the controlled networks, the generating specification is known, so we can directly test whether \textsc{Forge} recovers the correct terms. Some of the largest gains, including those on Caltech36 and Faux Magnolia, come from cases where the best baseline still has a large structural mismatch. Since $q(M)$ records the largest GOF discrepancy, these reductions represent substantial improvement over the baseline, not necessarily near-perfect absolute fit. \textsc{Forge} also often selects richer specifications, but PBIC improves on 11 networks, suggesting that the added terms usually improve fit enough to offset the complexity penalty.

The LLM and estimator comparisons show where tool choice matters in \textsc{Forge}. Across LLMs, effect directions are interpreted relatively consistently, while exact references to ERGM terms vary more. Estimator choice mainly affects fitting time and scale: SA makes the 16-network benchmark practical, while EE further reduces fitting time on Wiki-Vote. Wiki-Vote also shows an important limit: faster estimation does not make a poor specification fit well. Overall, \textsc{Forge} automates iterative ERGM specification while keeping model selection tied to statistical evaluation.

\section{Conclusion}

We introduced \textsc{Forge}, an agentic system for ERGM specification that combines LLM-based proposal with statistical fitting, diagnostics, and refinement, and returns a human-readable interpretation of the selected model. Across 16 benchmark networks, \textsc{Forge} improves on the best baseline on 13 networks and matches it on the other three, with a median 24.8\% improvement in goodness of fit. On 60 controlled networks with known generating specifications, \textsc{Forge} achieves exact specification recovery in 78.3\% of cases and reaches 90.8\% term-level F1. Together, these results show that \textsc{Forge} can turn the manual propose--fit--diagnose--refine cycle of ERGM specification into a systematic and reproducible agentic workflow while keeping model decisions grounded in statistical evidence.

\section*{Limitations}

\textsc{Forge} currently focuses on simple, static, binary networks without self-loops. Its search is bounded by the available ERGM terms, estimator settings, and simulation budget. Extending the system to longitudinal, multilayer, or bipartite networks would require the corresponding ERGM terms and fitting procedures.

Our experiments use a fixed goodness-of-fit procedure across 16 benchmark networks, and results may vary with the term set, LLM, estimator, or simulation budget. Absolute goodness of fit also remains challenging on some networks despite substantial relative improvements. Stage~IV explains associations represented by the fitted ERGM rather than causal effects. Its evaluation tests whether explanations preserve the supplied terms and coefficient directions, not whether they are useful to non-expert users.

\section*{Data and Code Availability}

All benchmark datasets are publicly available from the sources listed in the supplement. We will release the prompts, processed data, evaluation code, and outputs needed to reproduce the results.

\section*{Ethical Considerations}

The benchmark includes organizational, adolescent social, and security-related networks. Our analysis concerns network structure rather than individual intent, and should not be used for individual-level decisions without appropriate privacy and domain review. LLM-generated suggestions may also contain errors or biases, particularly when applied to private or sensitive data.

\bibliographystyle{ACM-Reference-Format}
\bibliography{references}

\end{document}